\documentclass[12pt]{llncs}
\usepackage[top=1.1in, bottom=1.1in, left=1.3in, right=1.3in]{geometry}
\usepackage{graphicx}
\usepackage{amsmath}
\usepackage{amssymb}
\usepackage{afterpage}
\usepackage{txfonts}

\def\Proof{\par\noindent{\bf Proof:}\indent}
\def\QED{\hfill$\Box$\par\vskip1em}
\def\PROG#1{$\mathcal{#1}$}

\begin{document}
\title{Packet Efficient Implementation of \\the Omega Failure Detector}

\author{Quentin Bramas\inst{1} \and Dianne Foreback\inst{2} \and 
        Mikhail Nesterenko\inst{2} \and
        S\'{e}bastien Tixeuil\inst{1,3}}
\institute{Sorbonne Universit\'{e}és, UPMC Univ Paris 06, CNRS, LIP6
  UMR 7606, 4 place Jussieu 75005 Paris \and Kent State University
  \and Institut Universitaire de France}

\maketitle

\begin{abstract}
We assume that a message may be delivered by packets through multiple
hops and investigate the feasibility and efficiency of an
implementation of the Omega Failure Detector under such an assumption.
To motivate the study, we prove that the existence and sustainability
of a leader is exponentially more probable in a multi-hop Omega
implementation than in a single-hop one.

An implementation is: \emph{message efficient} if all but finitely
many messages are sent by a single process; \emph{packet efficient} if
the number of packets used to transmit a message in all but finitely
many messages is linear w.r.t the number of processes, packets of
different messages may potentially use different channels, thus the
number of used channels is not limited; \emph{super packet efficient}
if the number of channels used by packets to transmit all but finitely
many messages is linear.

We present the following results for deterministic algorithms. If
reliability and timeliness of one message does not correlate with
another, i.e., there are no channel reliability properties, then a
packet efficient implementation of Omega is impossible. If eventually
timely and fair-lossy channels are considered, we establish necessary
and sufficient conditions for the existence of a message and packet
efficient implementation of Omega.  We also prove that the eventuality
of timeliness of channels makes a super packet efficient
implementation of Omega impossible.  On the constructive side, we
present and prove correct a deterministic packet efficient
implementation of Omega that matches the necessary conditions we
established.
\end{abstract}

\section{Introduction} 

%
%
The asynchronous system model places no assumptions on message
propagation delay or relative process speeds. This makes the model
attractive for distributed algorithm research as the results obtained
in the model are applicable to an arbitrary network and computer
architecture.  However, the fully asynchronous system model is not
well suited for fault tolerance studies. An elementary problem of
consensus, where processes have to agree on a single value, is
unsolvable even if only one process may crash~\cite{flp}: the
asynchrony of the model precludes processes from differentiating a
crashed and a slow process.

A failure detector~\cite{CT96} is a construct that enables the
solution to consensus or related problems in the asynchronous system
model.  Potentially, a failure detector may be very powerful and,
therefore, hide the solution to the problem within its
specification. Conversely, the weakest failure detector specifies the
least amount of synchrony required to implement
consensus~\cite{CHT96}. One such detector is Omega\footnote{In
  literature, the detector is usually denoted by the Greek
  letter. However, we use the letter to denote low complexity
  bound. To avoid confusion, we spell out the name of the failure
  detector in English.}.

Naturally, a failure detector may not be implemented in the
asynchronous model itself. Hence, a lot of research is focused on
providing the implementation of a detector, especially Omega, in the
least restrictive communication model. These restrictions deal with
timeliness and reliability of message delivery.  Aguilera et
al.~\cite{ADFT08} provide a remarkable Omega implementation which
requires only a single process to have eventually timely channels to
the other processes and a single process to have so called fair-lossy
channels to and from all other processes. Aguilera et al. present what
they call an efficient implementation where only a single process
sends infinitely many messages. In their work, Aguilera et
al. consider a direct channel as the sole means of message delivery
from one process to another. In this paper, we consider a more general
setting where a message may arrive to a particular process through
several intermediate processes. Otherwise, we preserve model
assumptions of Aguilera et al.

\ \\
\noindent\textbf{Our contribution.}  We study Omega implementation under
the assumption that a message may come to its destination through
other processes. 

To motivate this multi-hop Omega implementation approach, we consider
a fixed probability of channel timeliness and study the probability of
leader existence in a classic single-hop and in multi-hop
implementations.  We prove that the probability of leader existence
tends to zero for single-hop implementations and to one for multi-hop
ones as network size grows. Moreover, probability of leader persisting
while the timeliness of channel changes tends to zero for single-hop
and to infinity for multi-hop implementations.

If we consider deterministic algorithms, we study three classes of
Omega implementations: message efficient, packet efficient and super
packet efficient. In a message efficient implementation all but
finitely many messages are sent by a single process.  In a packet
efficient implementation, the number of packets in all but finitely
many transmitted messages is linear w.r.t. the number of processes in
the network. However, in a (simple) packet efficient implementation,
packets of different messages may use different channels such that
potentially all channels in the system are periodically used.  In a
super packet efficient implementation, the number of channels used in
all but finitely many messages is also linear w.r.t. to the number of
processes.

Our major results are as follows.  If timeliness of one message does
not correlate with the timeliness of another, i.e., there are no
timely channels, we prove that any implementation of Omega has to send
infinitely many messages whose number of packets is quadratic w.r.t to
the number of processes in the network. This precludes a packet
efficient implementation of Omega.  If eventually timely and
fair-lossy channels are allowed, we establish the necessary and
sufficient conditions for the existence of a packet efficient
implementation of Omega. We then prove that this eventuality of timely
and channels precludes the existence of a super packet efficient
implementation of Omega.  We present an algorithm that uses these
necessary conditions provides a message and packet efficient
implementation of Omega

\ \\
\noindent\textbf{Related work.}  The implementation of failure detectors
is a well-researched
area~\cite{AR10,BW09,DDFL10,HMSZ09,LLSC15,LFA04,MOZ05,MMR03,MRT06,PLL00}. Refer
to~\cite{ADFT08,AR10} for detailed comparisons of work related to the
kind of Omega implementation we are proposing. We are limiting our
literature review to the most recent and closest to ours studies.

Delporte-Gallet et al.~\cite{DDFL10} describe algorithms for
recognizing timely channel graphs. Their algorithms are super packet
efficient and may potentially be used to implement Omega. However,
their solutions assume non-constant size messages and perpetually
reliable channels. That is Delporte-Gallet et al. deviate from the
model of Aguilera et al. and the algorithms of Delporte-Gallet et
al. do not operate correctly under fair-lossy and eventually timely
channel assumptions.

A number of papers consider Omega implementation under various
modifications of Aguilera et al model.  Hutle et al.~\cite{HMSZ09}
implement Omega assuming a send-to-all message transmission primitive
where $f$ processes are guaranteed to receive the message
timely. Fernandez and Raynal~\cite{AR10} assume a process that is able
to timely deliver its message to a quorum of processes over direct
channels. This quorum and channels may change with each message. A
similar rotating set of timely channels is used by Malkhi et
al.~\cite{MOZ05}. Larrea et al.~\cite{LFA04} give an efficient
implementation of Omega but assume that all channels are eventually
timely. In their Omega implementation, Mostefaoui et al.~\cite{MMR03}
rely on a particular order of message interleaving rather than on
timeliness of messages. Biely and Widder~\cite{BW09} consider
message-driven (i.e., non-timer based) model and provide an efficient
Omega implementation.

There are several recent papers on timely solutions to problems
related to Omega implementation. Charron-Bost et al.~\cite{CBFN14} use
a timely spanning tree to solve approximate consensus. Lafuente et
al.~\cite{LLSC15} implement eventually perfect failure detector using
a timely cycle of processes.

\section{Notation and Definitions} 
\noindent
\textbf{Model specifics.} To simplify the presentation, we use an even
more general model than what is used in Aguilera et
al.~\cite{ADFT08}. The major differences are as follows.  We use
infinite capacity non-FIFO channels rather than single packet capacity
channels. Our channel construct makes us explicitly state the packet
fairness propagation assumptions that are somewhat obscured by the
single capacity channels.  

In addition, we do not differentiate between a slow process and a slow
channel since slow channels may simulate both.  Omega implementation
code is expressed in terms of guarded commands, rather than the usual
more procedural description. The operation of the algorithm is a
computation which is a sequence of these command executions. We
express timeouts directly in terms of computation steps rather than
abstract or concrete time. This simplifies reasoning about them.

Despite the differences, the models are close enough such that all of
the results in this paper are immediately applicable to the
traditional Omega implementation model.

\ \\
\noindent\textbf{Processes and computations.} A computer network
consists of a set $N$ of processes. The cardinality of this set is
$n$.  Each process has a unique identifier from $0$ to $n-1$.
Processes interact by passing messages through non-FIFO unbounded
communication channels.  Each process has a channel to all other
processes.  That is, the network is fully connected.  A message is
\emph{constant size} if the data it carries is in $O(\log n)$. For
example, a constant size message may carry several process identifiers
but not a complete network spanning tree.

Each process has variables and actions. The action has a \emph{guard}:
a predicate over the local variables and incoming channels of the
process. An action is enabled if its guard evaluates to \textbf{true}.
A \emph{computation} is a potentially infinite sequence of global
network states such that each subsequent state is obtained by
executing an action enabled in the previous state. This execution is a
computation \emph{step}.  Processes may crash. \emph{Crashed} process
stops executing its actions. \emph{Correct} process does not crash.

\ \\
\noindent\textbf{Messages and packets.}  We distinguish between a
packet and a message. \emph{Message} is particular content to be
distributed to processes in the network. \emph{Origin} is the process
that initiates the message. The identifier of the origin is included
in the message.  Messages are sent via packets. \emph{Packet} is a
portion of data transmitted over a particular channel. A message is
the payload of a packet.  A process may receive a packet and either
forward the message it contains or not.  A process may not modify it:
if a process needs to send additional information, the process may
send a separate message.  A process may forward the same message at
most once.  In effect, a message is transmitted to processes of the
network using packets. A particular process may receive a message
either directly from the origin, or indirectly possibly through
multiple hops.

\ \\
\noindent\textbf{Scheduling and fairness.} We express process
synchronization in terms of an adversarial scheduler. The scheduler
restrictions are as follows.  We do not distinguish slow processes and
slow packet propagation. A scheduler may express these phenomena
through scheduling process action execution in a particular way.  A
packet transmission immediately enables the packet receipt action in
the recipient process. A packet is lost if the receipt action is never
executed. A packet is not lost if it is eventually received.

\ \\
\noindent\textbf{Timers.} Timer is a construct with the following
properties. A timer can be reset, stopped and increased. It can also
be checked whether the timer is on or off.  It has a \emph{timeout
  integer} value and a \emph{timeout action} associated with it. A
timer is either a receiver timer or a sender timer.  If a
\emph{sender timer} is on, timeout action is executed once the
computation has at most the timeout integer steps without executing
the timer reset. If a \emph{receiver timer} is on, the timeout action
is executed once the computation has at least the timeout integer
steps without executing the timer reset.  Increasing the timer, adds
an arbitrary positive integer value to the timeout integer. An off
timer can be set to on by resetting it.

\ \\
\noindent\textbf{Reliable and timely messages and packets.}  A packet
is \emph{reliable} if it is received. A message is reliable if it is
received by every correct process. A channel is reliable if every
packet transmitted over this channel is reliable.
%
%
%

A channel is \emph{fair-lossy} if it has the following properties. If
there is an infinite number of packet transmissions over a particular
fair-lossy channel of a particular message type and origin, then
infinitely many are received. We assume that a fair-lossy channel is
not type discriminating. That is, if it is fair-lossy for one type and
origin, it is also fair-lossy for every pair of message type and
origin.

Observe that if there is an infinite number of message transmissions
of a particular message type and origin over a path that is
fair-lossy, then infinitely many succeed. There converse is true as
well: if there is an infinite number of successful message
transmissions, there must be a fair-lossy path between the origin an
the destination.

A packet is \emph{timely} if it is received within a bounded number of
computation steps.  Specifically, there is a finite integer $B$ such
that the packet is received within $B$ steps.  Naturally, a timely
packet is a reliable packet.  A message is timely if it is received by
every process via a path of timely packets.  A channel is timely if
every packet transmitted over this channel is timely.  A channel is
eventually timely if the number of non-timely packets it transmits is
finite. Note that a channel that transmits a finite number of packets
is always eventually timely.

The timely channel definition is relatively clear. The opposite,
non-timely channel, is a bit more involved. A channel that
occasionally delays or misses a few packets is not non-timely as the
algorithm may just ignore the missed packets with a large enough
timeout. Hence, the following definition. 

A channel is \emph{strongly non-timely} if the following holds.  If
there is an infinite number of packet transmissions of a particular
type and origin over a particular non-timely channel, then, for any
fixed integer, there are infinitely many computation segments of this
length such that none of the packets are delivered inside any of the
segments.

Similarly, the non-timeliness has to be preserved across multiple
channels, a message may not gain timeliness by finding a parallel
timely path, then, for example, the two paths may alternate delivering
timely messages. Therefore, we add an additional condition for
non-timeliness.

All paths between a pair of processes $x$ and $y$ are \emph{strongly
  non-timely} if $x$ sends an infinite number of messages to $y$, yet
regardless of how the message is forwarded or what path it takes, for
any fixed integer, there are infinitely many computation segments of
this length such that none of the messages are delivered inside any of
the segments.  Unless otherwise noted, when we discuss non-timely
channels and paths, we mean strongly non-timely channels and paths.

\ \\
\noindent\textbf{Communication models.} To make it easier to address
the variety of possible communication restrictions, we define several
models.  \emph{The dependable (channel) model} allows eventually or
perpetually reliable timely or fair-lossy channels. In the dependable
model, an algorithm may potentially discover the dependable channels
by observing packet propagation.  \emph{The general propagation model}
does not allow either reliable or timely channels.  Thus, one message
propagation is not related to another message propagation.

\ \\
\noindent\textbf{Message propagation graph.} \emph{Message propagation
  graph} is a directed graph over network processes and channels that
determines whether packet propagation over a particular channel would
be successful. This graph is connected and has a single source: the
origin process. This concept is a way to reason about scheduling of
the packets of a particular message.

Each message has two propagation graphs. In \emph{reliable propagation
 graph} $R$, each edge indicates whether the packet is received if
transmitted over this channel. In \emph{timely propagation graph} $T$
each edge indicates whether the packet is timely if transmitted over
this channel. Since a timely packet is a reliable packet, for the same
message, the timely propagation graph is a subgraph of the reliable
propagation graph.  In general, a propagation graph for each message
is unique. That is, even for the same source process, the graphs for
two messages may differ. This indicates that different messages may
take divergent routes.

If a channel from process $x$ to process $y$ is reliable, then edge
$(x,y)$ is present in the reliable propagation graph for every message
where process $x$ is present. In other words, if the message reaches
$x$ and $x$ sends it to $y$, then $y$ receives it.  A similar discussion
applies to a timely channel and corresponding edges in timely
propagation graphs.

Propagation graphs are determined by the scheduler in advance of the
message transmission. That is, the recipient process, depending on the
algorithm, may or may not forward the received message along a
particular outgoing channel. However, if the process forwards the
message, the presence of an edge in the propagation graph determines
the success of the message transmission.  Note that the process
forwards a particular message at most once. Hence, the propagation
graph captures the complete possible message propagation pattern.  A
process may crash during message transmission. This crash does not
alter propagation graphs.

\begin{proposition}\label{propRconnected}
A message is reliable only if its reliable propagation graph $R$
is such that every correct process is reachable from the origin
through non-crashed processes.
\end{proposition}

\begin{proposition}\label{propTconnected}
A message is timely only if its timely propagation graph $T$ is such
that every correct process is reachable from the origin through
non-crashed processes.
\end{proposition}

\ \\
\noindent\textbf{Omega implementation and its efficiency.}  An algorithm
that implements the Omega Failure Detector (or just Omega) is such
that in a suffix of every computation, each correct process outputs
the identifier of the same correct process.  This process is the
\emph{leader}.

An implementation of Omega is \emph{message efficient} if the origin
of all but finitely many messages is a single correct process and all
but finitely many messages are constant size. An implementation of
Omega is \emph{packet efficient} if all but finitely many messages are
transmitted using $O(n)$ packets.

An omega implementation is \emph{super packet efficient} if it is
packet efficient and the packets of all but finitely many messages are
using the same channels. In other words, if a packet of message $m_1$
is forwarded over some channel, then a packet of another message $m_2$
is also forwarded over this channel. The intent of a super packet
efficient algorithm is to only use a limited number of channels
infinitely often. Since a packet efficient algorithm uses $O(n)$
packets infinitely often, a super packet efficient algorithm uses
$O(n)$ channels infinitely often.

\section{Probabilistic Properties}
\label{secMotivation}
In this section, we contrast a multi-hop implementation of Omega and a
classic single-hop, also called direct channel, implementation.  We
assume each network channel is timely with probability $p$. The
timeliness probability of one channel is independent of this probability
of any other channel.

\ \\
\noindent\textbf{Leader existence probability.} We assume that the
leader may exist only if there is a process that has timely paths to
all processes in the network. In case of direct channel
implementation, the length of each such path must be exactly one.

As $n$ grows, Omega implementations behave radically differently.
Theorems~\ref{trmLeaderOneHop} and~\ref{trmLeaderMultiHop} state the
necessary conditions for leader existence and indicate that the
probability of leader existence for direct channel implementation
approaches zero exponentially quickly, while this probability for
multi-hop implementation approaches one exponentially quickly.  In
practical terms, a multi-hop omega implementation is far more likely
to succeed in establishing the leader.

\begin{theorem} \label{trmLeaderOneHop}
If the probability of each channel to be timely is $p < 1$, then the
probability of leader existence in any direct channel Omega
implementation approaches zero exponentially fast as $n$ grows.
\end{theorem}

\Proof Let $D_x$ be the probability that some process $x$ does not have direct
timely channels to all processes in the network. This probability is
$\mathbb{P}(D_x) = 1-p^{n-1}$.  For two distinct processes $x$ and $y$,
$D_x$ and $D_y$ are disjoint since channels are oriented. Thus, if $p
< 1$, the probability that no leader exists is
$\mathbb{P}(\bigcap_{x\in V}D_x) =
(1-p^{n-1})^{n}\overset{n\rightarrow + \infty}{\rightarrow} 1$. \QED

\begin{theorem} \label{trmLeaderMultiHop}
If the probability of each channel to be timely is $p < 1$, then the
probability of leader existence in any multi-hop Omega implementation
approaches $1$ exponentially fast as $n$ grows.
\end{theorem}

\Proof A channel is \emph{bitimely} if it is timely in both
directions.  The probability that there exists at least one process
such that there exist timely paths from this process to all other
processes is greater than the probability to reach them through
bitimely paths.  We use the probability of the latter as a lower bound
for our result.  If $p$ is the probability of a channel to be timely,
$\tilde{p} = p^2$ is the probability that it is bitimely.  Consider
graph $G$ where the edges represent bitimely channels. It is an
Erdos-Renyi graph where an edge exists with probability $\tilde{p}$.
It was shown (see~\cite{gilbert}) that $\mathbb{P}(G\textrm{ is
  connected}) \sim 1-n(1-\tilde{p})^{n-1}\overset{n\rightarrow +
  \infty}{\rightarrow} 1$. \QED

\ \\
\noindent\textbf{Leader stability.} As in previous subsection, we
assume the leader has timely paths to all other processes in the
network. If channel timeliness changes, this process may not have
timely paths to all other processes anymore. \emph{Leader stability
  time} is the expected number of rounds of such channel timeliness
change where a particular process remains the leader.

Again, direct channel and multi-hop implementations of Omega behave
differently. Direct channel leader stability time approaches zero as
$n$ increases and cannot be limited from below by fixing a particular
value of channel timeliness probability. Multi-hop leader stability
goes to infinity exponentially quickly. In a practical setting, a
leader is significantly more stable in a multi-hop Omega
implementation than in a direct channel one.

\begin{theorem}\label{trmStabilityDirect}
In any direct channel Omega implementation, if the probability of each
channel to be timely is $p < 1$, leader stability time goes
exponentially fast to $0$ as $n$ grows. If leader stability time is to
remain above a fixed constant $E > 0$, then the channel timeliness
probability $p$ must converge to $1$ exponentially fast as $n$ grows.
\end{theorem}

\Proof At a given time, a given process has timely channels to all
other processes with probability $p^{n-1}$. The number of rounds $X$ a
given process retains timely paths to all other processes follows a
geometric distribution $\mathcal{P}(X=r)=q^{r}(1-q)$, where $q =
p^{n-1}$. Thus, the expected number of rounds a process retains timely
channels to all other processes is $\frac{q}{1-q} =
\frac{p^{n-1}}{1-p^{n-1}}\sim p^{n-1}$, which tends exponentially fast
toward $0$ if $p$ is a constant less than $1$.

Assume $\mathbb{E}(X)$ converges towards a given fixed number $E$ as
$n$ tends towards infinity. That is, we need
$\lim_{n\rightarrow\infty} \mathbb{P}(G\textrm{ is connected}) =
\frac{1}{E+1}$. Then, $p^{n-1}$ tends to $\frac{1}{E+2}$, which
implies that $p$ converges towards $1$ exponentially fast. \QED

\begin{theorem}\label{trmStabilityMultiHop}
In any multi-hop Omega implementation, if the probability of each
channel to be timely is $p < 1$, leader stability time goes to
infinity exponentially fast as $n$ grows. If leader stability time is
to remain above a fixed constant $E > 0$, then channel timeliness
probability may converge to $0$ exponentially fast as $n$ grows.
\end{theorem}

\Proof If we fix $\tilde{p}$, $0<\tilde{p}<1$, we have
$\mathbb{P}(G\textrm{ is connected}) \sim 1-n(1-\tilde{p})^{n-1}$
(see~\cite{gilbert}). Then, the expected number of rounds a given
process retains timely paths to all other processes is asymptotically
$n^{-1}\left(\frac{1}{1-\tilde{p}}\right)^{n}$, which increases
exponentially fast.

Assume $\mathbb{E}(X)$ converges towards a given fixed number $E$ as
$n$ tends to infinity. This means that $$\lim_{n\rightarrow\infty}
\mathbb{P}(G\textrm{ is connected}) = \frac{1}{E+1} = e^{-e^{-c}}$$
Using well-known results of random graph theory~\cite{bollobas}, we
can take $$\tilde{p}(n) = \frac{\ln{n}}{n} + \frac{c}{n}=
\frac{\ln{n}}{n} - \frac{\ln\ln{(1+E)}}{n}$$ \QED

\section{Necessity and Sufficiency Properties}\label{secNecessary}

We now explore the properties of deterministic Omega implementation.

\ \\ \textbf{Model independent properties.} The below Omega
implementation properties are applicable to both general propagation
and dependable channel model.

\begin{theorem}\label{trmTimely}
In an implementation of Omega, at least one correct process needs to
send infinitely many timely messages.
\end{theorem} 

\Proof Assume \PROG{A} is an implementation of Omega where
every correct process sends a finite number of timely messages. Start
with a network where all but two processes $x$ and $y$ crash, wait
till all timely messages are sent. Since \PROG{A} is an implementation
of Omega, eventually $x$ and $y$ need to agree on the leader. Let it
be $x$.  Since all timely messages are sent, the remaining messages
may be delayed arbitrarily.  If $x$ now crashes, process $y$ must
eventually elect itself the leader. Instead, we delay messages from $x$
to $y$. The crash and the delay are indistinguishable to $y$ so it
elects itself the leader. We now deliver messages in an arbitrary
manner. Again, since \PROG{A} implements Omega, $x$ and $y$ should
agree on the leader. Let it be $y$. The argument for $x$ is similar.
We then delay messages from $y$ to $x$ forcing $x$ to select itself
the leader. We continue this procedure indefinitely. The resultant
sequence is a computation of \PROG{A}. However at least one process,
either $x$ or $y$, oscillates in its leader selection infinitely many
times. To put another way, this process never elects the leader. This
means that, contrary to the initial assumption, \PROG{A} is not an
implementation of Omega. This proves the theorem.  \QED

If single process sends an infinite number of messages in a message
efficient implementation of Omega, this process must be the
leader. Otherwise processes are not able to recognize the crash of the
leader. Hence, the corollary of Theorem~\ref{trmTimely}.

\begin{corollary}\label{corLeaderSends}
In a message efficient implementation of Omega, the leader must send
infinitely many timely messages.
\end{corollary}

\ \\
\noindent\textbf{General propagation model properties.} 

\begin{lemma} \label{lemNeedAll}
To timely deliver a message in the general propagation model, each
recipient process needs to send it across every outgoing channel,
except for possibly the channels leading to the origin and the sender.
\end{lemma}

\Proof Assume the opposite. There exists an algorithm \PROG{A} that
timely delivers message $m$ from the origin $x$ to all processes in
the network such that some process $y$ receives it timely yet does not
forward it to some process $z \neq x$.

Consider the propagation graph $T$ for $m$ to be as follows. 
 \[x \rightarrow y \rightarrow z \rightarrow \text{rest of the processes}\]
That is, the timely paths to all processes lead from $x$ to $y$ then to
$z$.  If \PROG{A} is such that $x$ sends $m$ to $y$, then, by
assumption, $y$ does not forward $m$ to $z$. Therefore, no process
except for $y$ gets $m$ through timely packets. By definition of the
timely message, $m$ is not timely received by these processes. If $x$
does not send $m$ to $y$, then none of the processes receive a timely
message. In either case, contrary to the initial assumption, \PROG{A}
does not timely deliver $m$ to all processes in the network.  \QED

The below corollary follows from Lemma~\ref{lemNeedAll}.

\begin{corollary}\label{corN2}
It requires $\Omega(n^2)$ packets to timely deliver a message in
the general propagation model.
\end{corollary}

Combining Corollary~\ref{corN2} and Theorem~\ref{trmTimely} we obtain
Corollary~\ref{corNoEff}.

\begin{corollary}\label{corNoEff}
In the general propagation model, there does not exist a message and
packet efficient implementation of Omega.
\end{corollary}

\begin{proposition} 
There exists a message efficient implementation of Omega in the
general propagation model where each correct process can send reliable
messages to the leader.
\end{proposition}

The algorithm that proves the above proposition is a straightforward
extension of the second algorithm in Aguilera et al.~\cite{ADFT08}
where every process re-sends received messages to every outgoing
channel.

\ \\
\noindent\textbf{Dependable channel model properties.}

\begin{lemma}\label{lemNeedFairlossy}
In any message efficient implementation of Omega, each correct process
must have a fair-lossy path to the leader.
\end{lemma}

\Proof Assume there is a message-efficient implementation \PROG{A} of
Omega where there is a correct process $x$ that does not have a
fair-lossy path to the leader. According to
Corollary~\ref{corLeaderSends}, $x$ itself may not be elected the
leader. Assume there is a computation $\sigma_1$ of \PROG{A} where
process $y \neq x$ is elected the leader. Note that fair-lossy
channels are not type discriminating. That is, if $x$ does not have a
fair-lossy path to $y$, but has a fair lossy path to some other
process $z$, then $z$ does not have a fair-lossy path to $y$ either.
Thus, there must be a set of processes $S \subset N$ such that $x \in
S$ and $y \notin S$ that do not have fair-lossy paths to processes
outside $S$.

Since \PROG{A} is message efficient, processes of $S$ only send a
finite number of messages to $y$.  Consider another computation
$\sigma_2$ which shares prefix with $\sigma_2$ up to the point were
the last message from processes of $S$ is received outside of
$S$. After that, all messages from $y$ to processes in $S$ and all
messages from $S$ to outside are lost. That is in $\sigma_2$, $y$ does
not have timely, or every fair-lossy, paths to processes of $S$. It is
possible that some other process $w$ is capable of timely
communication to all processes in the network. However, since \PROG{A}
is efficient, no other processes but $y$ is supposed to send
infinitely many messages.

Since all messages from $S$ are lost, $\sigma_1$ and $\sigma_2$ are
indistinguishable for the correct processes outside $S$.  Therefore,
they elect $y$ as the leader.  However, processes in $S$ receive no
messages from $y$. Therefore, they have to elect some other process
$u$ to be the leader. This means that \PROG{A} allows correct
processes to output different leaders. That is, \PROG{A} is not an
implementation of Omega.  \QED

We define a \emph{source} to be a process that does not have incoming
timely channels.

\begin{lemma}\label{lemNeedAllSources}
To timely deliver a message in the dependable channel model, each
recipient needs to send it across every outgoing channel to a source,
except for possibly the channels leading to the origin and the sender.
\end{lemma}

The proof of the above lemma is similar to the proof of
Lemma~\ref{lemNeedAll}. Observe that Lemma~\ref{lemNeedAllSources}
states that the timely delivery of a packet requires $n$ messages per
source. If the number of sources is proportional to the number of
processes in the network, we obtain the following corollary.

\begin{corollary}\label{corN2sources}
It requires $\Omega(n^2)$ packets to timely deliver a message in the
dependable channel model where the number of sources is proportional
to $n$.
\end{corollary}

\begin{theorem} \label{trmDependableChannel}
In the dependable channel model, the following conditions are
necessary and sufficient for the existence of a packet and message
efficient implementation of Omega: (i) there is at least one process
$l$ that has an eventually timely path to every correct process (ii)
every correct process has a fair-lossy path to $l$.
\end{theorem}

\Proof We demonstrate sufficiency by presenting, in the next section,
an algorithm that implements Omega in the dependable channel model
with the conditions of the theorem.

We now focus on proving necessity.  Let us address the first condition
of the theorem.  Assume there is a message and packet efficient
implementation \PROG{A} of Omega in the dependable channel model even
though no process has eventually timely paths to every correct
process. Let there be a computation of \PROG{A} where some process $x$
is elected the leader even though $x$ does not have a timely path to
each correct process.  According to Corollary~\ref{corLeaderSends},
$x$ needs to send infinitely many timely messages. According to
Corollary~\ref{corN2sources}, each such message requires $\Omega(n^2)$
packets. That is, \PROG{A} may not be message and packet
efficient. This proves the first condition of the theorem.  The second
condition immediately follows from Lemma~\ref{lemNeedFairlossy}.
\QED

The below theorem shows that (plain) efficiency is all that can be
achieved with the necessary conditions of
Theorem~\ref{trmDependableChannel}. That is, even if these conditions
are satisfied, supper packet efficiency is not possible.

\begin{theorem}\label{trmNotEventually}
There does not exist a message and super packet efficient
implementation of Omega in the dependable communication model even if
there is a process $l$ with an eventually timely path to every correct
process and every correct process has a fair-lossy path to $l$.
\end{theorem}

\Proof Assume the opposite. Suppose there exists a super packet
efficient algorithm \PROG{A} that implements Omega in the network
where some process $l$ has an eventually timely path to all correct
processes and every correct process has fair-lossy paths to $l$.

\begin{figure}
\centering
\includegraphics[width=6cm]{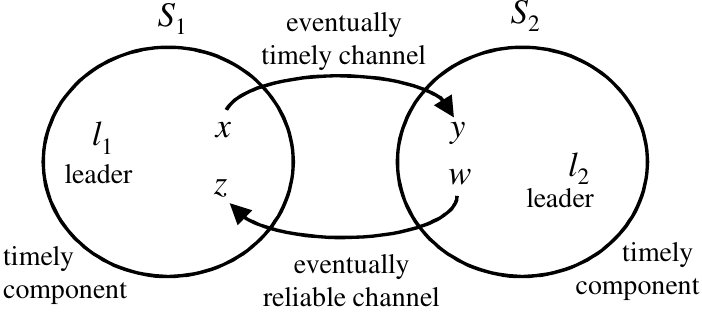}
\caption{Network for $\sigma_3$ computation of 
Theorem~\ref{trmNotEventually}.}\label{figNoEventual}
\end{figure}

Without loss of generality, assume the number of processes in the
network is even. Divide the processes into two sets $S_1$ and $S_2$
such that the cardinality of both sets is $n/2$. Refer to
Figure~\ref{figNoEventual} for illustration.  $S_1$ is completely
connected by timely channels. Similarly, $S_2$ is also completely
connected by timely channels. The dependability of channels between
$S_1$ and $S_2$ is immaterial at this point.

Consider a computation $\sigma_1$ of \PROG{A} on this network where
all processes in $S_1$ are correct and all processes in $S_2$ crashed
in the beginning of the computation. Since \PROG{A} is an
implementation of Omega, one process $l_1 \in S_1$ is elected the
leader. Since \PROG{A} is message efficient, only $l_1$ sends messages
infinitely often. Since \PROG{A} is super packet efficient, only
$O(n)$ channels carry theses messages infinitely often. Since the
network is completely connected, there are $(n/2)^2$ channels leading
from $S_1$ to $S_2$. This is in $O(n^2)$. Thus, there is least one
channel $(x,y)$ such that $x \in S_1$ and $y \in S_2$ that does not
carry messages from $l_1$ infinitely often.

Let us consider a computation $\sigma_2$ of \PROG{A} where all
processes $S_2$ are correct and all processes in $S_1$ crash in the
beginning of the computation.  Similar to $\sigma_1$, there is a
process $l_2 \in S_2$ that is elected the leader in $\sigma_2$, and
there is a channel $(z, w)$ such that $z \in S_2$ and $w \in S_1$ that
carries only finitely many messages of $l_2$.

We construct a computation $\sigma_3$ of \PROG{A} as follows. All
processes are correct. Channel dependability inside $S_1$ and $S_2$ is
as described above. All channels between $S_1$ and $S_2$ are
completely lossy, i.e., they lose every transmitted message. An
exception is channel $(x,y)$ that becomes timely as soon as it loses
the last message it is supposed to transmit. Similarly, channel
$(z,w)$ becomes reliable as soon as it loses the last message.

To construct $\sigma_3$, we interleave the actions of $\sigma_1$ and
$\sigma_2$ in an arbitrary manner.  Observe that to processes in $S_1$
computations $\sigma_1$ and $\sigma_3$ are
indistinguishable. Similarly, to processes in $S_2$, the computations
$\sigma_2$ and $\sigma_3$ are indistinguishable.

Let us examine the constructed computation closely. Sets $S_1$ and
$S_2$ are completely connected by timely channels, and $(x,y)$,
connecting $S_1$ and $S_2$ is eventually timely. This means that $l_1$
has an eventually timely path to every correct process in the
network. Moreover, due to channel $(z,w)$, every process has a
fair-lossy path to $l_1$. That is, the conditions of the theorem are
satisfied.  However, the processes of $S_1$ elect $l_1$ as their
leader while the processes of $S_2$ elect $l_2$. This means that the
processes do not agree on the single leader. That is, contrary to the
initial assumption, \PROG{A} is not an implementation of Omega. The
theorem follows.  \QED

\section{\PROG{MPO}: Message and Packet Efficient Implementation of Omega}
In this section we present an algorithm we call
\PROG{MPO} that implements Omega in the fair-lossy channel
communication model. As per Theorem~\ref{trmDependableChannel}, we
assume that there is at least one process that has an eventually timely
path to every correct process in the network and every correct
process has a fair-lossy path to this process.

\afterpage{
\thispagestyle{empty}
\begin{figure}
\scriptsize
\begin{tabbing}
12\=1234\=1234\=1234\=1234\=1234\=1234\=1234\=123\kill
\>\textbf{constants}\\
\>\> $p$ // process identifier \\
\>\>  $N$ // set of network process identifiers, cardinality is $n$ \\
\>\>  $timers[p]$ length is $TO$ \\
\>\textbf{variables} \\
\>\> $leader$, initially $\bot$ // local leader \\
\>\> $phases [n]$, initially zero // current phase number \\
\>\> $edges [n][n]$, initially zero // edge fault weights \\
\>\> $arbs [n]$, initially arbitrary // arborescences \\
\>\> $timers [n]$, initially $timers[p]$ on, others off \\
\>\>\>length of $timers[x]: x\neq p$ is arbitrary // timer to send/receive a message \\
\>\> $shout$, initially zero // process id to send $alive$ to all neighbors\\
\ \\
\>\textbf{actions}\\
\>\>$timeout(timers[q]) \longrightarrow$ \\
\>\>\>$\textbf{if}\ p = q\ \textbf{then}$ // own/sender timeout, compute arb rooted in $p$ based on $edges$\\ 
\>\>\>\>$newArb = arborescence(edges, p)$ \\
\>\>\>\>$newLeader := minWeight((arbs[r]: r \neq p:  on(timers[r])), newArb)) $\\
\>\>\>\>$\textbf{if}\ leader \neq newLeader\ \textbf{then}$ // leadership changes \\
\>\>\>\>\>$\textbf{if}\ newLeader = p\ \textbf{then}$ // $p$ gains leadership\\
\>\>\>\>\>\>$arbs[p] := newArb$ \\ 
\>\>\>\>\>\>$\textbf{send}\ startPhase(p, phases[p], arbs[p])\ \textbf{to}\ N/{p}$ \\
\>\>\>\>\>$\textbf{if}\ leader = p\ \textbf{then}$ // $p$ loses leadership\\
\>\>\>\>\>\>$phases[p] := phases[p]+1$ \\
\>\>\>\>\>\>$\textbf{send}\ stopPhase(p, phases[p])\ \textbf{to}\ N/{p}$ \\
\>\>\>\>\>$leader := newLeader$ \\ 
\>\>\>\>$\textbf{else}$ // leadership persists \\
\>\>\>\>\>$\textbf{if}\ leader = p\ \textbf{then}$ \\
\>\>\>\>\>\>$shout := shout + 1\ \textbf{mod}\ n$\\
\>\>\>\>\>\>$\textbf{if}\ shout \neq p\ \textbf{then}$ \\
\>\>\>\>\>\>\>$\textbf{send}\ alive(p, phases[p], shout)\ \textbf{to}\ arbs[p](p.children)$ \\
\>\>\>\>\>\>$\textbf{else}$ // my turn to shout\\
\>\>\>\>\>\>\>$\textbf{send}\ alive(p, phases[p], shout)\ \textbf{to}\ N/{p}$ \\
\>\>\>\>$reset(timers[p])$ // own timer never off\\
\>\>\>$\textbf{else}$ // neighbor timeout/receiver timeout, assume failed, increase, do not reset\\
\>\>\>\>$\textbf{send}\ failed(q, p, arbs[q](p.parent))\ \textbf{to}\ N/{p}$ \\
\>\>\>\>$increase(timers[q])$ \\
\ \\
\>\> $\textbf{receive}\ startPhase(q, phase, arb)$ for the first time $\longrightarrow$ \\
\>\>\>// if new phase, propagate message, reset timer\\ 
\>\>\>$\textbf{if}\ p \neq q \wedge phases[q] \leq phase\ \textbf{then}$ \\
                       
\>\>\>\>$arbs[q] := arb$ \\
\>\>\>\>$phases[q] := phase$ \\
\>\>\>\>$\textbf{send}\ startPhase(q, phase, arb)\ \textbf{to}\ N/{p}$ \\ 
\>\>\>\>$reset(timers[q])$ \\ 
\ \\
\>\>$\textbf{receive}\ stopPhase(q, phase)$ for the first time $\longrightarrow$  \\
\>\>\> $\textbf{if}\ p \neq q \wedge phase[q] < phase\ \textbf{then}$ \\
\>\>\>\>$phases[q] := phase$ \\
\>\>\>\>$\textbf{send}\ stopPhase(q, phase)\ \textbf{to}\ N/{p}$ \\
\>\>\>\>$stop(timers[q])$\\
\ \\
\>\> $\textbf{receive}\ alive(q, phase, sh)$ for the first time 
$\textbf{from}\ r \ \longrightarrow$  \\
\>\>\>$\textbf{if}\ p \neq q \wedge phase[q] = phase\ \textbf{then}$ \\
\>\>\>\>$\textbf{if}\ r = arbs[q](p.parent)\ \textbf{then}$ // received through arborescence\\
\>\>\>\>\>$\textbf{if}\ sh \neq p\ \textbf{then}$ \\
\>\>\>\>\>\>$\textbf{send}\ alive(q, phase, sh)\ \textbf{to}\ arbs[q](p.children)$ \\
\>\>\>\>\>$\textbf{else}$ // my turn to shout\\
\>\>\>\>\>\>$\textbf{send}\ alive(q, phase, sh)\ \textbf{to}\ N/{p}$ \\
\>\>\>\>\>$reset(timers[q])$ \\ 
\>\>\>\>$\textbf{else}$ // received from elsewhere \\
\>\>\>\>\>$\textbf{if}\ \mathit{off}(timers[q])\ \textbf{then}$ \\
\>\>\>\>\>\>$reset(timers[q])$ \\
\ \\
\>\> $\textbf{receive}\ failed(q, r, s)$ for the first time $\longrightarrow$ \\
\>\>\>$\textbf{if}\ p = q\ \textbf{then}$ // if $p$'s \emph{alive} failed\\ 
\>\>\>\>$edges[s][r] := edges[s][r] + 1$ // increase weight of edge from parent\\
\>\>\>$\textbf{else}$ \\
\>\>\>\>$\textbf{send}\ failed(q, r, s)\ \textbf{to}\ N/{p}$ \\ 
\end{tabbing}
\caption {Message and packet efficient implementation of Omega
  \PROG{MPO}.}\label{figAlg}
\end{figure}
\clearpage
}

\ \\
\noindent\textbf{Algorithm outline.} The code of the algorithm is shown in
Figure~\ref{figAlg}.  
The main idea of \PROG{MPO} is for processes to
attempt to claim the leadership of the network while discovering the
reliability of its channels. Each process weighs each channel by the
number of messages that fail to come across it. The lighter channel is
considered more reliable. If a process determines that it has the
lightest paths to all processes in the network, the process tries to
claim leadership of the network.

The leadership is obtained in phases. First, the leader candidate
sends \emph{startPhase} message. Then, the candidate periodically
sends \emph{alive} message. In case an \emph{alive} fails to reach one
of the processes on time, the recipient replies with \emph{failed}.
The size of \emph{startPhase} depends on the network size. The size of
the other message types is constant.

The routes of the messages vary. Messages that are only sent finitely
many times are \emph{broadcast}: sent across every channel in the
network. Once one process receives such a message for the first time,
the process re-sends it along all of its outgoing channels.
Specifically, \emph{startPhase}, \emph{stopPhase} and \emph{failed}
are broadcast. The leader sends \emph{alive} infinitely often. Hence,
for the algorithm to be packet efficient, \emph{alive} has to be sent
only along selected channels. Message \emph{alive} is routed through
the channels that the origin believes to be the most reliable.

Specifically, \emph{alive} is routed along the channels of a minimum
weight \emph{arborescence}: a directed tree rooted in the origin
reaching all other processes. The arborescence is computed by the
origin once it claims leadership. It is sent in the \emph{startPhase}
that starts a phase. Once each process receives the arborescence, the
process stores it in the $arbs$ array element for the corresponding
origin. After receiving \emph{alive} from a particular origin, the
recipient consults the respective arborescence and forwards the
message to the channels stated there.

In addition to routing \emph{alive} along the arborescence, each
process takes turns sending the leader's \emph{alive} to all its
neighbors.  The reason for this is rather subtle: see
Theorem~\ref{trmNotEventually} for details.  Due to crashes and
message losses, $arbs$ for the leader at various processes may not
reach every correct process.  For example, it may lead to a crashed
process. Thus, some processes may potentially not receive \emph{alive}
and, therefore, not send \emph{failed}. Since \emph{failed} are not
sent, the leader may not be able to distinguish such a state from a
state with correct $arbs$.

To ensure that every process receives \emph{alive}, each process, in
turn, sends \emph{alive} to its every neighbor rather than along most
reliable channels.  Since only a single process sends to all neighbors
a particular \emph{alive} message, the packet complexity remains
$O(n)$.

Message \emph{failed} is sent if a process does not receive a timely
\emph{alive}. This message carries the parent of the process which was
supposed to send the \emph{alive}. That is, the sender of
\emph{failed} blames the immediate ancestor in the arborescence.  Once
the origin of the missing \emph{alive}, receives \emph{failed}, it
increments the weight of the appropriate edge in $edges$ that stores
the weights of all channels.  If a process has timely outgoing paths
to all processes in the network, its arborescence in $edges$
convergences to these paths.


\ \\
\noindent\textbf{Action specifics.} The algorithm is organized in five
actions.  The first is a timeout action, the other four are
message-receipt actions.

The timeout action handles two types of timers: sender and
receiver. Process $p$'s own timer ($q=p$) is a sender timer. It is
rather involved. This timer is always on since the process resets it
after processing. First, the process computes the minimum weight of
the arborescence for each leader candidate. A process is considered a
leader candidate if its timer is on.  Note that since $p$'s own timer
is always on, it is always considered.

The process with the minimum weight arborescence is the
new leader. If the leadership changes ($leader \neq newLeader$),
further selection is made. If $p$ gains leadership ($newLeader = p$),
then $p$ starts a new phase by updating its own minimum-weight
arborescence and broadcasting \emph{startPhase}. If $p$ loses
leadership, it increments its phase and broadcasts \emph{stopPhase}
bearing the new phase number.

If the leadership persists ($leader = newLeader$) and $p$ is the
leader, it sends \emph{alive}. Process $p$ keeps track of whose turn
it is to send \emph{alive} to all its neighbors in the \emph{shout}
variable. The variable's value rotates among the ids of all
processes in the network.

The neighbor timer ($q \neq p$) is a receiver timer.  If the
process does not get \emph{alive} on time from $q$, then $p$ 
sends \emph{failed}. In case the process sends \emph{failed}, it also
increases the timeout value for the timer of $q$ thus attempting to
estimate the channel delay.

For our algorithm, the timer integers are as follows. The sender timer
is an arbitrary constant integer value $TO$. This value controls how
often \emph{alive} is sent. It does not affect the correctness of the
algorithm. Receiver timers initially hold an arbitrary value. The
timer integer is increased every time there is a timeout. Thus, for an
eventually timely channel, the process is able to estimate the
propagation delay and set the timer integer large enough that the
timeout does not occur. For untimely channels, the timeout value may
increase without bound.

The next four actions are message receipt handling. Note that a single
process may receive packets carrying the same message multiple times
across different paths. However, every process handles the message at
most once: when it encounters it for the first time. Later duplicate
packets are discarded.

The second action is \emph{startPhase} handling. The process copies
the arborescence and phase carried by the message, rebroadcasts it and
then resets the \emph{alive} receiver timer associated with the origin
process. The third action is the receipt of \emph{stopPhase} which
causes the recipient to stop the appropriate timer.

The forth action is \emph{alive} handling.  If \emph{alive} is the
matching phase, it is further considered. If \emph{alive} comes
through the origin's arborescence, the receiver sends \emph{alive} to
its children in the origin's arborescence or broadcasts it.  The
process then resets the timer to wait for the next \emph{alive}. If
\emph{alive} comes from elsewhere, that is, it was the sender's turn
to send \emph{alive} to all its neighbors, then $p$ just resets the
timeout and waits for an \emph{alive} to arrive from the proper
channel.  This forces the process to send \emph{failed} if
\emph{alive} does not arrive from the channel of the arborescence.

The last action is \emph{failed} handling. If \emph{failed} is in
response to an \emph{alive} originated by this process ($p=q$) then
the origin process increments the weight of the edge from the parent
of the reporting process to the process itself according to the
message arborescence.  If \emph{failed} is not destined to this
process, $p$ rebroadcasts it.

\section{\PROG{MPO} Correctness Proof}
\noindent\textbf{Correctness proof definitions.}  Throughout this
section, $l$ is the identifier of the process that has eventually
timely paths to all other processes. For simplicity, we assume that
$l$ is the single such process.  Denote $B$ as the maximum number of
steps in any timely channel propagation delay.  Process $p$ is a
\emph{local leader} if $leader_p=p$, i.e., the process elected itself
the leader. A process may be a local leader but not the global
leader. That is, several processes may be local leaders in the same
state.  Let $realArbs(x)$ for the origin process $x$ be the relation
defined by $arbs[x](y.children)$ at every process $y$. That is,
$realArbs(x)$ is the distributed relation that determines how $alive$
messages are routed if they are originated by $x$.

%
%
%
\begin{lemma} \label{manyFailed}
For any local leader process $x$ and another correct process $y$ such
that $y$ is not reachable from $x$ through timely channels over
correct processes in $realArbs(x)$, either (i) $realArbs(x)$ changes
or (ii) $x$ loses leadership, changes phase or receives infinitely
many \emph{failed} messages.
\end{lemma}

\Proof To prove the lemma, it is sufficient to show that if
$realArbs(x)$ does not change and $x$ does not lose the leadership or
change phase, then $x$ receives infinitely many \emph{failed}.

Let $S$ be a set of correct processes that are reachable from $x$
through timely channels and through correct processes in
$realArbs(x)$. Since $y$ is not reachable from $x$, $S \neq N$.
Recall that every process has fair-lossy paths to all processes in the
network. Therefore, there is such a path from $x$ to $y$.  This means
that there is a process $z \in S$ such that it has a fair-lossy
channel to $w \notin S$.

Let us examine process $w$ closer.  The network is completely
connected. Therefore, all other processes from $S$ have channels to
$w$. Note that at least one channel, from $z$ is fair-lossy. Moreover,
since $w$ does not belong to $S$, if $realArbs(x)$ reaches $w$, the
path to $w$ is not timely.

Since $x$ is a local leader and does not lose its leadership, it sends
infinitely many \emph{alive} messages. Other processes forward these
\emph{alive} along $realArbs(x)$.  Also, by the design of the
algorithm, every process takes turn sending \emph{alive} to all of its
neighbors rather than forwarding it along $realArbs(x)$. Let us
examine the receipt of these messages by $w$.

Process $z$ belongs to $S$.  That is, the path from $x$ to $z$ in
$realArbs(x)$ is timely. This means that it receives and sends
infinitely many \emph{alive} originated by $x$. Since the channel from
$z$ to $w$ is fair-lossy, infinitely many of these \emph{alive} are
delivered to $w$. In addition, $w$ possibly receives \emph{alive} from
other processes of $S$.  Since, none of these channels are part of
$realArbs(x)$, when $w$ receives \emph{alive} from processes in $S$,
it resets the corresponding receive timer only when the timer is
off. The timer is turned off only when the timeout is executed and
\emph{failed} is broadcast.

The only possible way this receive timer is reset without the timeout
action execution is when $w$ receives \emph{alive} through
$realArbs(x)$. However, the path from $x$ to $w$ in $realArbs(x)$ is
not timely.  By the definition of non-timely paths, there are
infinitely many computation segments of arbitrary fixed length where
no \emph{alive} from $x$ is delivered to $w$.  This means that,
regardless of the timeout variable value at $w$, the \emph{alive}
messages generate receiver timeouts.  That is, infinitely many
timeouts are executed at $w$. Each timeout generates a \emph{failed}
message broadcast by $w$. Since there are infinitely many broadcasts,
infinitely many succeed in reaching $x$. Hence, the lemma.  \QED

\begin{lemma}\label{xGetsFailed}
If each process $x \neq l$ is a local leader in infinitely many states
then it receives infinitely many \emph{failed} messages.
\end{lemma}

\Proof Let $x \neq l$ be a local leader in infinitely many states of a
particular computation of the algorithm. Once a process assumes local
leadership, it may lose it either by (i) increasing the weight of its
minimum weight arborescence (ii) by recording an arborescence
$arbs[y]$ for a process $y$ with lower weight than $arbs[x]$.

A process increases the weight of its arborescence only when it gets a
\emph{failed} message. Thus, to prove the lemma we need to consider
the second case only.

Since $x$ is a local leader in infinitely many states, it must gain
local leadership back after losing it to another process $y$. By the
design of the algorithm, the weight of the arborescence of any process
in $arbs$ may only increase.  This means that once $x$ gains the
leadership back from $y$, $x$ may not lose it to $y$ again without
increasing the weight of its own minimum weight arborescence. Thus,
either $x$ increases the weight of its arborescence or, eventually, it
has the lightest arborescence among the leader candidates.

In case $x$ has the lightest arborescence, it either becomes heavier
than some other leader candidate's or $x$ gets infinitely many
\emph{failed}. However, only the latter part of the statement needs to
be proven since $x$ gains leadership infinitely often.

If $x$ is a local leader, it does not send \emph{startPhase} or
\emph{stopPhase}. Let us consider the state where all
\emph{startPhase} packets are delivered. In this case $realArbs(x)$
does not change. Since $x \neq l$, even if all correct processes are
reachable from $x$ in $realArbs(x)$, some links in $realArbs(x)$ are
not timely. Then, according to Lemma~\ref{manyFailed}, $x$ gets
infinitely many \emph{failed}.

To summarize, if $x \neq l$ is a local leader in infinitely many
states, it receives infinitely many \emph{failed}. \QED

\begin{lemma}\label{lLeaderOften}
Process $l$ is a local leader in infinitely many states.
\end{lemma}

\Proof According to Lemma~\ref{xGetsFailed} either each process $x
\neq l$ stops gaining local leadership or the weight of its minimum
arborescence grows infinitely high.  If the latter is the case, $x$
has to gain and lose local leadership infinitely often. In this case,
it sends \emph{startPhase} infinitely may times. Message
\emph{startPhase} is broadcast. Since every process $x$ has fair-lossy
paths to $l$, by the definition of fair-lossy paths, infinitely many
broadcasts succeed. This means that the weight of $arbs[x]$ at $l$
grows without bound. Therefore, if $l$ loses local leadership, it
gains it back infinitely often.  \QED

The below lemma follows immediately from the operation of the
algorithm.

\begin{lemma} \label{timersFixed}
The timer length of $timers[l]$ at every process either stops
increasing or it reaches $TO+B*(n-1)$
\end{lemma}

And the below lemma follows from the assumption that the leader
has an eventually timely path to every correct process.

\begin{lemma} \label{broadcastsOk}
In every computation, there is a suffix where each broadcast message
sent by $l$ is timely delivered to every correct process.
\end{lemma}


\begin{lemma}\label{timelyOk}
An edge leading to process $x$ in a timely path in $realArbs(l)$ at $l$
generates only finitely many \emph{failed}.
\end{lemma}

\Proof The origin starts every phase with \emph{startPhase}, then
periodically sends zero or more \emph{alive} and then possibly ends
the phase with a \emph{stopPhase} that carries the phase number
greater than \emph{alive} and \emph{startPhase}.

Message \emph{failed} is generated only when the timer expires at the
receiving process.  The timer is reset by \emph{startPhase} and
\emph{alive}. The timer is stopped by \emph{stopPhase}.

We prove the lemma by showing that the timer reset by messages of a
particular phase expires only finitely many times.  We start our
consideration from the point of the computation where the conditions
of Lemmas~\ref{timersFixed} and~\ref{broadcastsOk} hold.

Only \emph{alive} and \emph{startPhase} may reset the timeout. Since
the conditions of Lemma~\ref{broadcastsOk} hold, \emph{startPhase} is
delivered within $B(n-1)$ computation steps to all processes. Message
\emph{alive} may be received earlier than \emph{startPhase}. However,
since such \emph{alive} carries a phase number that differs from the
number stored at the recipient process, the message is ignored. If
\emph{alive} arrives after \emph{startPhase}, the reasoning is similar
to the case where \emph{alive} is sent after \emph{startPhase} which
is to be considered next.

Every \emph{alive} sent after \emph{startPhase} delivery, travels over
the timely path in $realArbs(l)$. At most every $TO$ number of steps,
either another \emph{alive} or \emph{stopPhase} is sent. Since the
path in $realArbs(l)$ is timely, \emph{alive} arrives at most after $TO +
B(n-1)$ steps. Due to Lemma~\ref{broadcastsOk}, the same is true of
\emph{stopPhase}. That is, after \emph{alive} is received, either
another \emph{alive} or \emph{stopPhase} is received within
$TO+B(n-1)$ steps. The receipt of \emph{alive} resets the timeout. The
receipt of \emph{stopPhase} stops it. Due to Lemma~\ref{timersFixed},
the timer does not expire.

Moreover, after the receipt of \emph{stopPhase}, the subsequent
\emph{alive} messages are ignored since \emph{stopPhase} carries a
greater phase number. That is, after \emph{stopPhase} is received, the
timer is never reset or expires due to the messages of this phase.
\QED

\begin{lemma}\label{untimelyGone}
Every untimely edge in $realArbs(l)$ leading to a correct process
either gets removed or $l$ gets infinitely many \emph{failed}
messages.
\end{lemma}

\Proof Due to Lemma~\ref{lLeaderOften}, process $l$ is a
local leader in infinitely many states. Through the argument similar
to that of Lemma~\ref{xGetsFailed}, we can show that eventually either
$l$ gets \emph{failed} and increases the weight of its minimum
arborescence or its minimum arborescence becomes the lightest among
the leader candidates. Then, $l$ can lose leadership only if it gets
\emph{failed}.

In this case, according to Lemma~\ref{manyFailed}, $l$ receives
infinitely many failed messages or either loses leadership, changes
phase or changes $realArbs(l)$. Observe that $l$ may change phase only
when it receives \emph{failed}. It loses leadership only if it gets
\emph{failed}. The change of $realArbs(l)$ happens only when $l$
broadcasts \emph{startPhase} after changing phase and, therefore,
getting \emph{failed}. Due to Lemma~\ref{lLeaderOften}, it gains the
leadership back infinitely often.

That is, in any case, as long as $realArbs(l)$ contains an untimely
edge leading to a correct process, $l$ gets infinitely many
\emph{failed}. \QED

The below lemma follows from Lemmas~\ref{timelyOk}
and~\ref{untimelyGone}.

\begin{lemma} \label{arbsTimely}
Every computation of \PROG{MPO} contains a suffix where each channel of
$realArbs(l)$ is timely.
\end{lemma}

\begin{lemma}\label{arbsSame}
Every computation of \PROG{MPO} contains a suffix where $realArbs(l)$
is the same as $arbs[l]$ in process $l$.
\end{lemma}

\Proof We start our consideration from the point where the conditions
of Lemma~\ref{arbsTimely} hold.  Suppose $realArbs(l)$ and $arbs[l]$ differ
for some process $x$. By the design of the algorithm, this may happen
only if $arbs[l]$ in process $x$ has an earlier phase than in
$l$. However, since phases differ, \emph{alive} sent by $l$ are
ignored by $x$. This leads to either $x$ sending \emph{fail} to $l$ or
claiming leadership. In either case, $l$ sends
\emph{startPhase}. According to Lemma~\ref{broadcastsOk}, this
broadcasts succeeds which synchronizes $arbs[l]$ and $realArbs(l)$.
\QED

\begin{theorem}\label{trmMPO}
Algorithm \PROG{MPO} is a message and packet efficient implementation
of Omega in the fair-lossy channel model.
\end{theorem}

\Proof First, we prove that \PROG{MPO} implements Omega.  Indeed,
lemma~\ref{lLeaderOften} shows that $l$ is a local leader in
infinitely many states. Lemmas~\ref{timelyOk} and~\ref{untimelyGone}
show that $l$ gets finitely many \emph{failed}.  According to
Lemma~\ref{xGetsFailed}, every process $x \neq l$ either stops being a
local leader or gets infinitely many \emph{failed}. This means that at
any process the arborescence of $l$ will eventually be lighter than
any other leader contender.

According to Lemma~\ref{lLeaderOften}, $l$ sends infinitely many
\emph{alive} messages along $realArbs(l)$. Due to
Lemma~\ref{untimelyGone}, $realArbs(l)$ eventually has no untimely
channels. Since $l$, according to Lemma~\ref{untimelyGone}, receives
only finitely many \emph{failed}, due to Lemma~\ref{manyFailed},
$realArbs(l)$ eventually has timely paths from $l$ to every correct
process. According to Lemma~\ref{arbsSame}, $realArbs(l)$ and
$arbs[l]$ are eventually the same.

This means that $l$ will be a leader contender in every correct
process. Since it has the lightest arborescence, it becomes the leader
at every correct process. In other words, \PROG{MPO} is a correct
implementation of Omega.

By the design of the algorithm, once $l$ has the lightest arborescence
and all correct processes drop out of leadership contention, $l$ is
the only process that sends \emph{alive} messages. By definition,
\PROG{MPO} is message efficient. 

The messages are routed along $arbs[l]$. It is an arborescence. Hence,
the number of such messages is in $O(n)$.  In addition, each process
takes a turn sending \emph{alive} to its neighbors. This is another
$O(n)$ packets. Therefore, the packet complexity of \PROG{MPO} is in
$O(n)$.  \QED

\section{Algorithm Extensions}

%
%

We conclude the paper with several observations about \PROG{MPO}.  The
algorithm trivially works in a non-completely connected network
provided that the rest of the assumptions used in the algorithm
design, such as eventually timely paths from the leader to all correct
processes, are satisfied. Similarly, the algorithm works correctly if
the channel reliability and timeliness is origin-related. That is, a
channel may be timely for some, not necessarily incident, process $x$,
but not for another process $y$.

Algorithm \PROG{MPO} may be modified to use only constant-size
messages.  The only non-constant size message is \emph{startPhase}.
However, the message type is supposed to be timely. So, instead of
sending a single large message, the modified \PROG{MPO} may instead
send a sequence of fixed-size messages with the content to be
re-assembled by the receivers. If one of the constituent messages does
not arrive on time, the whole large message is considered lost.



\newpage

\begin{thebibliography}{10}

\bibitem{ADFT08}
Marcos~Kawazoe Aguilera, Carole Delporte{-}Gallet, Hugues Fauconnier, and Sam
  Toueg.
\newblock On implementing omega in systems with weak reliability and synchrony
  assumptions.
\newblock {\em Distributed Computing}, 21(4):285--314, 2008.

\bibitem{AR10}
Antonio~Fern{\'{a}}ndez Anta and Michel Raynal.
\newblock From an asynchronous intermittent rotating star to an eventual
  leader.
\newblock {\em {IEEE} Trans. Parallel Distrib. Syst.}, 21(9):1290--1303, 2010.

\bibitem{BW09}
Martin Biely and Josef Widder.
\newblock Optimal message-driven implementations of omega with mute processes.
\newblock {\em ACM Transactions on Autonomous and Adaptive Systems (TAAS)},
  4(1):4, 2009.

\bibitem{bollobas}
B\'{e}la Bollob\'{a}s.
\newblock {\em Random graphs}.
\newblock Cambridge University Press, 2 edition, October 2001.

\bibitem{CHT96}
Tushar~Deepak Chandra, Vassos Hadzilacos, and Sam Toueg.
\newblock The weakest failure detector for solving consensus.
\newblock {\em Journal of ACM}, 43(4):685--722, 1996.

\bibitem{CT96}
Tushar~Deepak Chandra and Sam Toueg.
\newblock Unreliable failure detectors for reliable distributed systems.
\newblock {\em Journal of the ACM}, 43(2):225--267, 1996.

\bibitem{CBFN14}
Bernadette Charron-Bost, Matthias F{\"u}gger, and Thomas Nowak.
\newblock Approximate consensus in highly dynamic networks.
\newblock {\em arXiv preprint arXiv:1408.0620}, 2014.

\bibitem{DDFL10}
Carole Delporte{-}Gallet, St{\'{e}}phane Devismes, Hugues Fauconnier, and Mikel
  Larrea.
\newblock Algorithms for extracting timeliness graphs.
\newblock In {\em Structural Information and Communication Complexity, 17th
  International Colloquium, {SIROCCO} 2010, Sirince, Turkey, June 7-11, 2010.
  Proceedings}, pages 127--141, 2010.

\bibitem{flp}
Michael~J. Fischer, Nancy~A. Lynch, and Michael~S. Paterson.
\newblock Impossibility of distributed consensus with one faulty process.
\newblock {\em J. ACM}, 32(2):374--382, 1985.

\bibitem{gilbert}
E.~N. Gilbert.
\newblock Random graphs.
\newblock {\em Ann. Math. Statist.}, 30(4):1141--1144, 12 1959.

\bibitem{HMSZ09}
Martin Hutle, Dahlia Malkhi, Ulrich Schmid, and Lidong Zhou.
\newblock Chasing the weakest system model for implementing $\omega$ and
  consensus.
\newblock {\em {IEEE} Trans. Dependable Sec. Comput.}, 6(4):269--281, 2009.

\bibitem{LLSC15}
Alberto Lafuente, Mikel Larrea, Iratxe Soraluze, and Roberto Corti{\~n}as.
\newblock Communication-optimal eventually perfect failure detection in
  partially synchronous systems.
\newblock {\em Journal of Computer and System Sciences}, 81(2):383--397, 2015.

\bibitem{LFA04}
Mikel Larrea, Antonio Fern{\'{a}}ndez, and Sergio Ar{\'{e}}valo.
\newblock On the implementation of unreliable failure detectors in partially
  synchronous systems.
\newblock {\em {IEEE} Trans. Computers}, 53(7):815--828, 2004.

\bibitem{MOZ05}
Dahlia Malkhi, Florin Oprea, and Lidong Zhou.
\newblock $\omega$ meets paxos: Leader election and stability without eventual
  timely links.
\newblock In {\em Distributed Computing}, pages 199--213. Springer, 2005.

\bibitem{MMR03}
Achour Mostefaoui, Eric Mourgaya, and Michel Raynal.
\newblock Asynchronous implementation of failure detectors.
\newblock In {\em 2013 43rd Annual IEEE/IFIP International Conference on
  Dependable Systems and Networks (DSN)}, pages 351--351. IEEE Computer
  Society, 2003.

\bibitem{MRT06}
Achour Most{\'{e}}faoui, Michel Raynal, and Corentin Travers.
\newblock Time-free and timer-based assumptions can be combined to obtain
  eventual leadership.
\newblock {\em {IEEE} Trans. Parallel Distrib. Syst.}, 17(7):656--666, 2006.

\bibitem{PLL00}
Roberto~De Prisco, Butler~W. Lampson, and Nancy~A. Lynch.
\newblock Revisiting the {PAXOS} algorithm.
\newblock {\em Theor. Comput. Sci.}, 243(1-2):35--91, 2000.

\end{thebibliography}
\end{document}